\documentclass[twocolumn,aps,prl]{revtex4-2}
\usepackage{times}
\usepackage{amsmath,amsfonts,amssymb}
\usepackage{amsthm,mathrsfs}
\usepackage{soul,bm,array,graphicx,bbold,multirow}
\usepackage[normalem]{ulem}
\usepackage[makeroom]{cancel}
\usepackage[usenames,dvipsnames]{xcolor}
\usepackage[colorlinks=true,citecolor=blue,linkcolor=red]{hyperref}
\usepackage{pdfpages}
\makeatletter
\AtBeginDocument{\let\LS@rot\@undefined}
\makeatother
\newcolumntype{x}[1]{>{\centering\let\newline\\\arraybackslash\hspace{0pt}}p{#1}}

\DeclareMathAlphabet{\mathbbold}{U}{bbold}{m}{n}

\newcounter{subeqn} %
\renewcommand{\Re}{\operatorname{Re}}
\renewcommand{\Im}{\operatorname{Im}} 
\makeatletter
\@addtoreset{subeqn}{equation}
\makeatother

\renewcommand{\Re}{\operatorname{Re}}
\renewcommand{\Im}{\operatorname{Im}}

\begin{document}
	
\title{Complex Frequency Fingerprint: Interacting Driven Non-Hermitian Skin Effect}

\author{Zhesen Yang$^{1,2}$}
\email[Corresponding author: ]{yangzs@xmu.edu.cn}	
\author{Zihan Wang$^1$, Juntao Huang$^{1,3}$, Zijian Zheng$^1$}
\author{Jiangping Hu$^{4,5,6}$}
\email[Corresponding author: ]{jphu@iphy.ac.cn}

\affiliation{$^1$ Department of Physics, Xiamen University, Xiamen 361005, Fujian Province, China}
\affiliation{$^2$ Asia Paciﬁc Center for Theoretical Physics, Pohang 37673, Korea}
\affiliation{$^3$ Jiujiang Research Institute of Xiamen University, Jiujiang,332000, China}
\affiliation{$^4$ Beijing National Laboratory for Condensed Matter Physics and Institute of Physics, Chinese Academy of Sciences, Beijing 100190, China}
\affiliation{$^5$ School of Physical Sciences, University of Chinese Academy of Sciences, Beijing 100190, China}
\affiliation{$^6$ New Cornerstone Science Laboratory, Beijing, 100190, China}
	
\date{\today}
	
\begin{abstract}

The excitation properties of quantum many-body systems are encoded in their response functions. 
These functions define an associated response Hamiltonian, which is intrinsically non-Hermitian due to the dissipative nature of retarded responses, even in closed systems. 
By analyzing its eigenvalues and eigenstates, one obtains a unique characterization of the system, referred to as the complex frequency fingerprint.
Using this framework, we demonstrate that interactions alone can give rise to both point-gap topology and the non-Hermitian skin effect. 
Unlike the dissipation-induced skin effect, this interaction-driven phenomenon exhibits pronounced frequency dependence. 
We further introduce a complex-frequency density of states framework that distinctly separates non-Hermitian skin modes from topological edge modes. 

\end{abstract}
	
\maketitle

{\em\color{red}Introduction.}---The use of response functions to characterize quantum many-body systems has a long and fruitful history~\cite{Coleman_2015t,bookhenrik,fetter}. 
For example, spectroscopic techniques such as Angle-Resolved Photoemission Spectroscopy (ARPES)~\cite{Coleman_2015t,RevModPhys.93.025006,RevModPhys.75.473} probe the single-particle spectral function to map electronic band structures, while Scanning Tunneling Microscopy (STM)~\cite{Coleman_2015t,RevModPhys.59.615,stmbook} measures the local density of states to reveal atomic-scale features and impurity states. 
Similarly, transport measurements, including conductivity and the Hall effect~\cite{RevModPhys.82.3045,PhysRevLett.49.405}, effectively probe the current-current response function, providing crucial insights into quasiparticle dynamics and topological invariants.

However, upon closer inspection, these experimental and theoretical approaches are found to predominantly reveal local information.
Although powerful, measurements such as the local density of states or two-point correlation functions are inherently limited to real-space or momentum-space neighborhoods. 
As a result, the rich non-local information encoded in the full structure of the response function, such as its eigenvalues, eigenstates, and their frequency-dependent behavior, remains largely unexplored.
Therefore, as proposed in Ref.~\cite{CFF,CFD} , analyzing the global matrix structure of response functions represents a significant and promising frontier for diagnosing new phases of matter and achieving a more complete understanding of quantum many-body phenomena. 
Since the resulting response functions are inherently non-Hermitian, their eigenvalues are complex. 
These complex eigenvalues and their corresponding eigenstates constitute the complex frequency fingerprint (CFF)~\cite{CFF} of a quantum many-body system.
Nevertheless, since the full response function is difficult to compute for generic quantum many-body systems, it remains challenging to systematically apply this approach, particularly in characterizing its frequency-dependent behavior, in a practical and general way.

In this work, we employ a semiclassical approximation to investigate a time-reversal symmetry-broken Su–Schrieffer–Heeger (SSH) model incorporating an onsite, sublattice-resolved Hubbard interaction. 
By leveraging the CFF framework, we conduct a comprehensive spectral and eigenstate analysis of the single-particle response function, which leads to the discovery of an interesting phenomenon: an interaction-induced point-gap topology accompanied by a non-Hermitian skin effect (NHSE)~\cite{Prl1210868,PrlYSW,Prl123170401,songfeiprl,PhysRevB.108.235422,Prl121kunst,transferkunst,Prb971214,prlOkuma,PrlYM,prbhigherorder,secnhse,entangle-nhse,PrbKOM,kawabata2025hopfbifurcation,Prl125126402,PrlYZFH,Prl125186,Prlborgnia,llhcritical,lchTR,prlgjb,tpswitch,Shen_2025,PhysRevB.104.195102,skinclus,PhysRevLett.133.216601,qin2025,Qin_2025,PhysRevLett.132.096501,lyc,Prl129070401,PhysRevLett.129.013903,PhysRevB.110.094308,li2025solitontransitionsmediatedskinmode,PhysRevApplied.16.057001,tpswitch,PhysRevB.105.195131,PhysRevLett.134.243805,lpm2-vcb4,li2025dissipation,PhysRevB.111.205418,PhysRevResearch.4.033122,PhysRevB.106.064208,PhysRevB.107.L220205,PhysRevA.107.043315,PhysRevLett.132.086502,PhysRevLett.132.063804,jp,PhysRevB.105.125421,Yuce_2021,PhysRevB.109.094308,Komis:23,PhysRevA.103.043329,PhysRevA.106.053315,PhysRevB.107.045131,PhysRevB.108.245114,PhysRevB.108.155114,PhysRevB.105.L180401,PhysRevA.110.L020201}. 
This effect fundamentally extends the notion of NHSE into the realm of interacting systems, revealing how many-body correlations can engender topologically non-trivial spectral structures.

Notably, these emergent phenomena exhibit pronounced frequency dependence. 
For example, both the point-gap topology~\cite{PrxUeda,Prxkawabata,prlOkuma,PhysRevLett.123.066405,PhysRevB.105.165137,Ashida,NHTP-rev,Prlborgnia,Ghatak_2019,PhysRevB.106.L121102,Rmp93015005,Dingkrev,Lin2023Topo,FoaTorres,PhysRevB.99.235112,PhysRevB.99.125103,PhysRevB.106.205147,PhysRevB.111.064310,PhysRevB.107.075118} and the associated NHSE disappear when the driving frequency exceeds a certain characteristic regions. 
To quantitatively capture this frequency-dependent behavior, we introduce a novel diagnostic tool, the commutator norm between the free Hamiltonian and the interaction-induced self-energy. 
This commutator norm serves as a measure of non-commutativity between free-particle dynamics and correlation effects, and it exhibits qualitatively agreement with our numerical results, highlighting its utility in characterizing interaction-induced spectral topological transitions.

Beyond spectral topology, the CFF framework also enables the resolution of quasiparticle resonances across the complex frequency plane. 
By defining complex-frequency local density of states and tunneling density of states, we demonstrate that although non-Hermitian skin modes appear spatially localized at the boundaries, they fundamentally act as transmission modes rather than truly localized states. 
As a comparison, the topological edge modes are real localized modes, and have no contributions to the tunneling. 
This distinction provides further insight into the physical foundation of the NHSE.

{\em\color{red}Complex frequency fingerprint.}---We begin our discussion with a quantum many-body system, aiming to understand its corresponding single-particle physics.
The system is described by the Hamiltonian:
\begin{equation}
	\hat{H}=\hat{H}_0+\hat{H}_i=\sum_{ij}t_{ij}\hat{\psi}^{\dag}_i\hat{\psi}_j+\sum_i\frac{U_i}{2}\hat{\psi}^{\dag}_i\hat{\psi}^{\dag}_i\hat{\psi}_i\hat{\psi}_i,\label{Eq1}
\end{equation}
where $\hat{\psi_i}$ and its complex conjugate $\hat{\psi}^{\dag}_i$ are bosonic annihilation and creation operators, respectively, at site $i$. 
The index $i$ encodes both the unit cell and sublattice; for example, sites $i=1,2,3,4,...$ correspond to $1A, 1B, 2A, 2B,...$, respectively.

The non-interacting term, $\hat{H}_0$, describes an SSH model with broken time-reversal symmetry~\cite{Prl125186}. 
Under periodic boundary conditions (PBC), it can be expressed in momentum space as
\begin{equation}
	\hat{H}_0=\sum_k\hat{\Psi}^{\dag}_k \mathcal{H}_0(k) \hat{\Psi}_k,
\end{equation}
where $\hat{\Psi}^{\dag}_k=(\hat{\psi}^{\dag}_{k,A},\hat{\psi}^{\dag}_{k,B})$, $\hat{\Psi}_k=(\hat{\psi}_{k,A},\hat{\psi}_{k,B})^{\mathsf{T}}$, and
\begin{equation}
	\mathcal{H}_0(k) = (t_1 + t_2 \cos k) \sigma_x + t_2 \sin k \sigma_y + \lambda \sin k \sigma_z. \label{Eq2}
\end{equation}
Although the $\sigma_z$ term breaks time-reversal symmetry, it preserves inversion symmetry, i.e., $\sigma_x\mathcal{H}_0(k)\sigma_x=\mathcal{H}_0(-k)$. 
The interacting term, $H_i$, incorporates a sublattice-dependent on-site Hubbard interaction, with $U_i = U_A$ for odd $i$ (sublattice A) and $U_i = U_B$ for even $i$ (sublattice B).

To probe the nontrivial effects of interactions from a single-particle perspective, we employ the recently proposed complex frequency fingerprint (CFF) method~\cite{CFF}. 
The core of this approach is to define an effective single-particle response Hamiltonian, $H^{\rm Response}(\omega_0)$, based on the steady-state response function, which is labeled by $\chi_{\rm steady}(\omega_0)$:
\begin{equation}
	H^{\rm Response}(\omega_0)=\omega_0-[\chi_{\rm steady}(\omega_0)]^{-1},\label{E4}
\end{equation}
The method then involves tracing the evolution of its eigenvalues and eigenstates as the driving frequency $\omega_0$ is varied from $-\infty$ to $+\infty$.

Consequently, the primary task in applying the CFF is to compute the full matrix of the single-particle response function, $\chi_{\rm steady}(\omega_0)$, in the steady state.

{\em\color{red}Response function and response Hamiltonian.}---
We now detail the computation of the $ij$-th matrix element of the response function, i.e., $[\chi_{\rm steady}(\omega_0)]_{ij}$. 
Here, the indices $i$ and $j$ represent site and sublattice labels; for instance, $j=2$ corresponds to site $1B$.
The calculation involves three steps:

\begin{figure}[b]
	\centering
	\includegraphics[width=0.95\columnwidth]{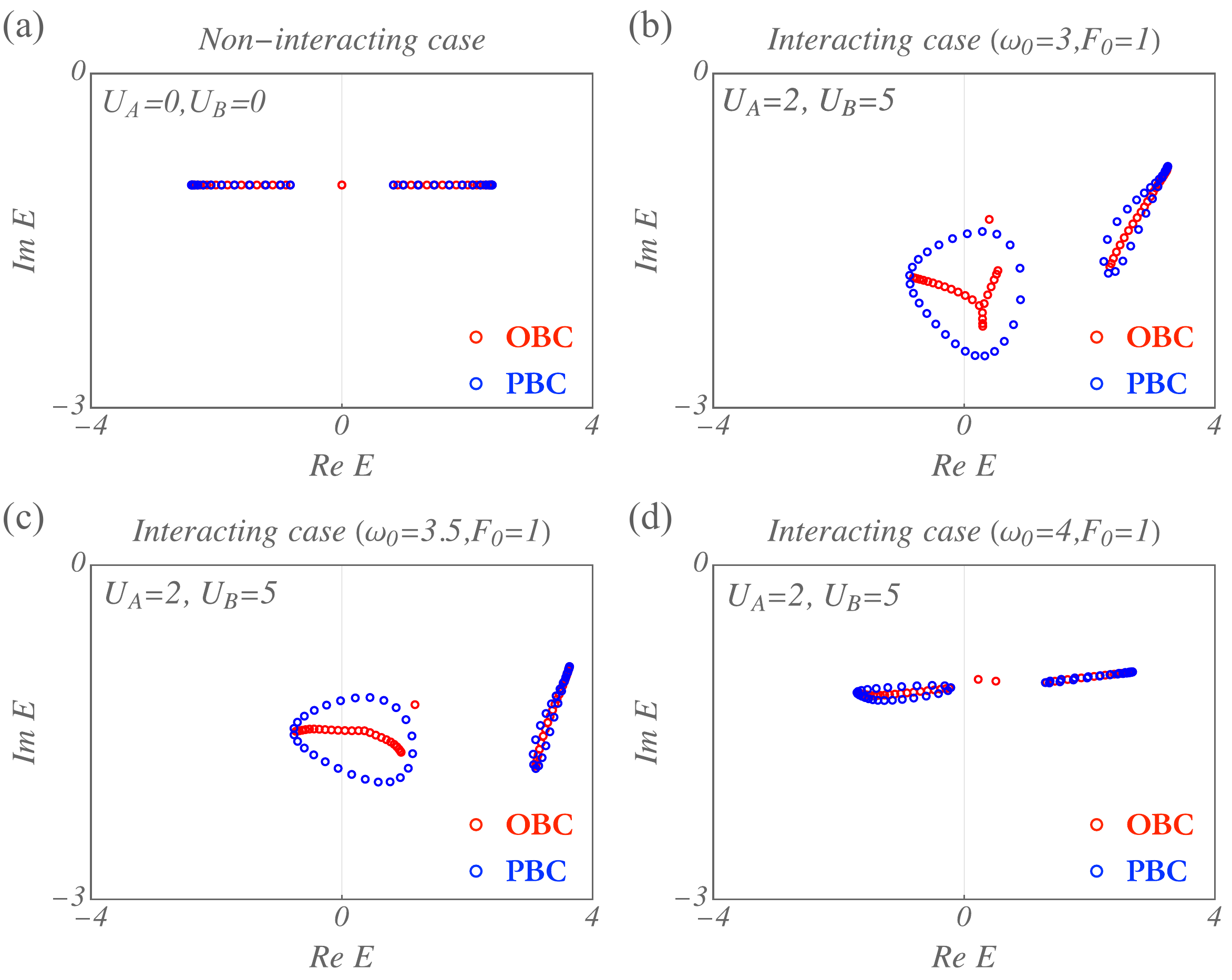}
	\caption{
		(a) Energy spectrum under periodic (PBC) and open (OBC) boundary conditions for the non-interacting case ($U_A=U_B=0$). The isolated eigenvalues near $\Re[E]=0$ are topological edge states.
		(b)-(d) PBC vs. OBC spectra for the interacting case with $F_0=1$ and $\omega_0=3,3.5,4$, respectively. A large point gap emerges for $\omega_0=3$, and the spectral discrepancy between PBC and OBC indicates the non-Hermitian skin effect. This point gap shrinks as $\omega_0$ increases and nearly vanishes at $\omega_0={\color{red}4}$, where the spectrum resembles the non-interacting limit.
	}
	\label{F1}
\end{figure}

Firstly, a single-particle drive is applied to the $j$th site by adding the following Hamiltonian to Eq.~\ref{Eq1}:
\begin{equation}
	\hat{H}^{\bm{F}_j}(t)=F_0e^{-i\omega_0t}\theta(t)\hat{\psi}^{\dag}_j + \text{h.c.}\label{Eq3}
\end{equation}
Here, $\bm{F}_j=(0,..., F_0,...,0)^{\mathsf{T}}$ is a vector whose only non-zero component corresponds to the driven site $j$, and $F_0$ and $\omega_0$ are the driving amplitude and frequency, respectively, $\theta(t)$ is the Heaviside step function. 


As detailed in the Appendix I, 
under the semiclassical approximation, the system's response is determined by the following non-linear differential equation:
\begin{equation}
	\left[ i\frac{\partial}{\partial t}-(\mathcal{H}_0-i\eta)-\mathcal{H}_i \right] \bm{\psi}^{\bm{F}_j}(t) = \bm{F}_je^{-i\omega_0 t}\theta(t).\label{Eq4}
\end{equation}
Here, $\mathcal{H}_0$ is the real-space representation of Eq.~\ref{Eq2}, $\eta > 0$ is a numerical convergence parameter, and $\mathcal{H}_i$ is the non-linear, diagonal operator:
\begin{equation}
	\mathcal{H}_i = \mathrm{diag}\left(U_A|\psi_{1}^{\bm{F}j}(t)|^2, U_B|\psi_{2}^{\bm{F}_j}(t)|^2,\ldots\right).\label{diaH}
\end{equation}
$\bm{\psi}^{\bm{F}_j}(t)=(\psi_1^{\bm{F}_j}(t),\psi_2^{\bm{F}_j}(t),... )$ is the time evolution of the mean value for $\hat{\bm{\psi}}^{\bm{F}_j}(t)$.
We use the initial condition $\bm{\psi}^{\bm{F}_j}(t=0)=0$ in our numerical calculations.

Finally, the matrix element of the response function is computed from the steady-state ratio of the response at site $i$ to the drive at site $j$:
\begin{equation}
	[\chi_{\rm steady}(\omega_0)]_{ij}=\lim_{t\rightarrow\infty}\frac{\psi_i^{\bm{F}_j}(t)}{F_0e^{-i\omega_0 t}}.
	\label{Eq6}
\end{equation}
By repeating this procedure for all source sites $j$, the full response matrix $\chi_{\rm steady}(\omega_0)$ is constructed numerically. 
We note that recent experiments~\cite{CFFcite1,CFFcite2,PhysRevLett.134.126603} have demonstrated the feasibility of this measurement protocol.

{\em\color{red}Interaction driven point gap topology.}---We now present the numerical results obtained using the method described above. 
The parameters for our simulations are set as follows: $t_1 = 0.8$, $t_2 = 1.6$, $\lambda=1$, $\eta = 1$, and a system size of $N=25$ unit cells.

As shown in Fig.~\ref{F1} (a), the non-interacting system under PBC exhibits zero spectral winding, while the spectrum under open boundary conditions (OBC) features two (degenerate) topological edge modes.
When interactions are introduced ($U_A=2,U_B=5$), both the PBC and OBC spectra are modified, as shown in Fig.~\ref{F1} (b)-\ref{F1} (d) for $\omega_0=3,3.5,4$ and $F_0=1$. 
Notably, a nonzero point gap emerges under PBC, and the corresponding OBC eigenvalues collapse into arcs, indicating the onset of the non-Hermitian skin effect~\cite{prlOkuma,Prl125126402}. 
Furthermore, both spectra exhibit a strong dependence on the driving frequency $\omega_0$.

These results can be understood within the self-energy formalism. From the response Hamiltonian, we define an associated single-particle self-energy:
\begin{equation}
	H^{\rm Response}(\omega_0)=\mathcal{H}_0-i\eta+\Sigma(\omega_0). 
\end{equation}
Here, $\Sigma(\omega_0)$ depends on the driving frequency $\omega_0$, amplitude $F_0$, and interaction strengths $U_A$ and $U_B$. 
As will be proved in the Appendix II, the self-energy in our model depends only on $\omega_0$ and the modified interaction $U_{\alpha}F_0^2$ ($\alpha=A,B$). 
We therefore fix $F_0=1$ throughout our discussion.

\begin{figure}[t] 
	\centering
	\includegraphics[width=1\columnwidth]{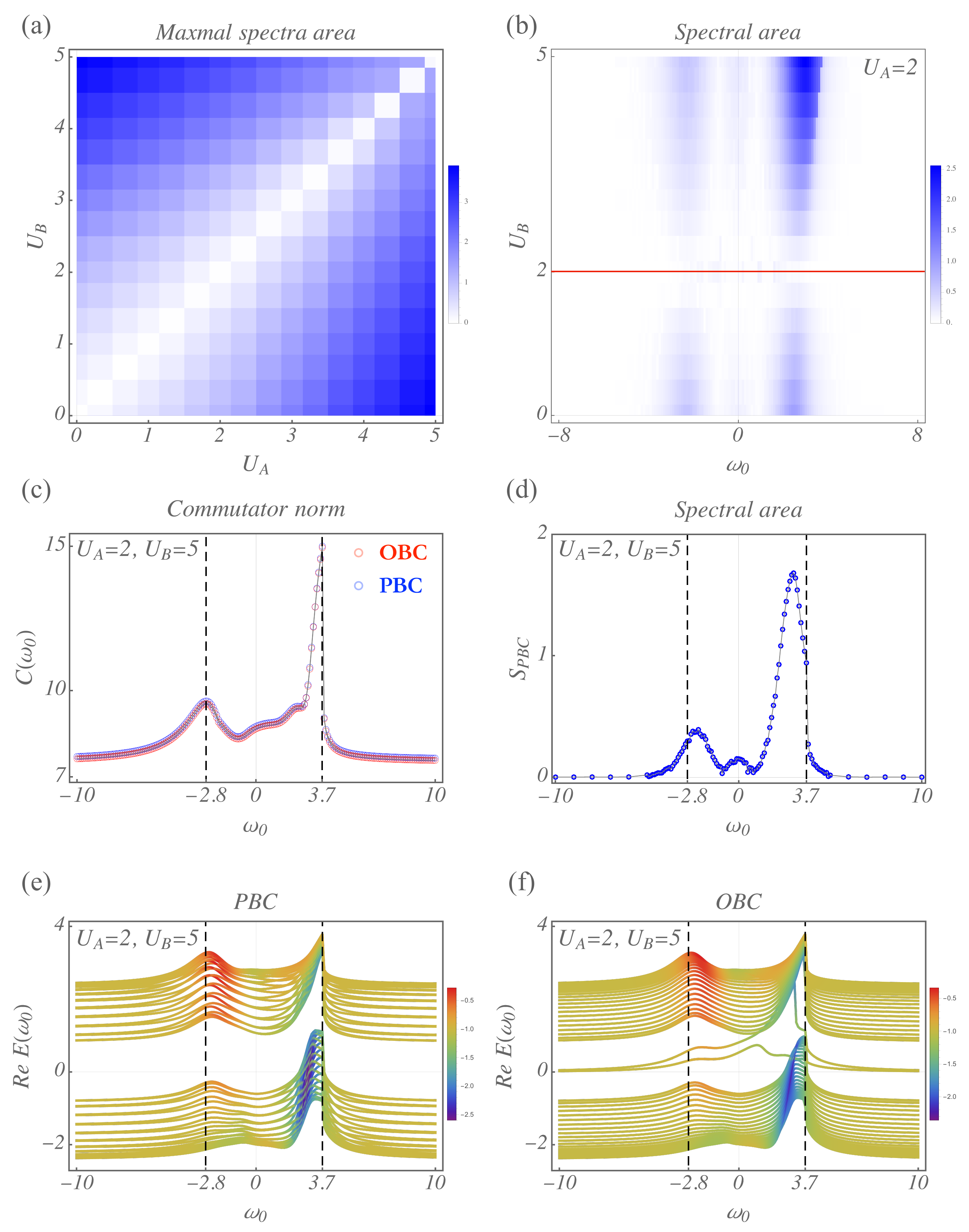}
	\caption{(a) The maximal area enclosed by the PBC spectra as a function of parameters \(U_A\) and \(U_B\), which vary from 0 to 5. A zero value indicates the absence of point-gap topology.  
	(b) PBC spectral area as a function of \(U_B\) and \(\omega_0\), with \(U_A\) fixed at 2. Two peaks are observed, corresponding to point-gap topology.  
	(c) Commutator norm \(C(\omega_0)\) under OBC (red) and PBC (blue) for \(U_A = 2\) and \(U_B = 5\), plotted as a function of driving frequency \(\omega_0\). The vertical dashed lines mark the two largest values.  
	(d) PBC spectral area as a function of \(\omega_0\) at \(U_A = 2\) and \(U_B = 5\), showing pronounced peaks at the characteristic frequencies identified in (c).  
	(e–f) Real part of eigenvalues \(\text{Re}[E_n(\omega_0)]\) versus \(\omega_0\) at \(U_A = 2\) and \(U_B = 5\) under (e) PBC and (f) OBC. The color scale represents the value of \(\text{Im}[E_n(\omega_0)]\).  }
	\label{F2}
\end{figure}

The self-energy $\Sigma(\omega_0)$ is generally a non-Hermitian matrix, which can induce point-gap topology and the associated non-Hermitian skin effect when inversion symmetry is broken, for example, $U_A \neq U_B$~\cite{Prl125186}.
This is confirmed by our numerical results. 
In Fig.~\ref{F2} (a), we plot the maximal spectral winding area over all real $\omega_0$ as a function of $U_A$ and $U_B$. 
The vanishing of this quantity indicates the absence of point-gap topology, which occurs precisely along the line $U_A = U_B$. 
Furthermore, Fig.~\ref{F2} (b) shows the spectral area versus $\omega_0$ and $U_B$ for a fixed $U_A=2$. 
It is clear that the spectral winding area is indeed zero for all $\omega_0$ along the line $U_B = U_A = 2$ (red line), confirming that point-gap topology and the skin effect cannot be induced when inversion symmetry is preserved.
Besides this, another interesting feature is that there are two resonance peaks over $\omega_0$ even $U_B$ is varied from $0$ to $5$.

To understand the origin of these two resonance peaks, we note that the condition $[\mathcal{H}_0, \Sigma(\omega_0)] = 0$ precludes the emergence of point gap topology and the non-Hermitian skin effect~\cite{Rmp93015005}. 
Let's take the OBC as an example to explain this point. 
When this condition is satisfied, the self-energy cannot alter the OBC eigenstates.
As a result, $\mathcal{H}_0+\Sigma(\omega_0)$ must have the same eigenbasis with $\mathcal{H}_0$.
This implies that the degree of non-commutativity between $\mathcal{H}_0$ and $\Sigma(\omega_0)$ provides a natural measure for the point gap size, which is manifested here as the two resonance peaks. 
Motivated by this, we define  a commutator norm:
\begin{equation}
	C(\omega_0) = ||[\mathcal{H}_0, \Sigma(\omega_0)]||,
\end{equation}
where $||A||$ is the spectral norm (i.e., the largest singular value) of matrix $A$. This norm vanishes, $C(\omega_0)=0$, if the operators commute at a frequency $\omega_0$.

Fig.~\ref{F2} (c) displays $C(\omega_0)$ for $U_A=2$ and $U_B=5$, with results for PBC and OBC shown in blue and red, respectively. 
Two dominant peaks are observed near $\omega_0 = -2.8$ and $3.7$. 
For comparison, Fig.~\ref{F2} (d) shows the area enclosed by the PBC spectrum for the same parameters. 
The two quantities exhibit qualitatively identical behavior, with only minor discrepancies in the peak positions. 
This agreement confirms that the commutator norm $C(\omega_0)$ successfully captures the fundamental physics of our model: 
it quantifies how strongly the self-energy $\Sigma(\omega_0)$ modifies the eigenstates of the non-interacting Hamiltonian $\mathcal{H}_0$. 
This is in stark contrast to conventional single-band Fermi liquids, where $[\mathcal{H}_0(k), \Sigma(k,\omega)]$ vanishes across broad regions near the Fermi surface, resulting in a vanishing $C(\omega_0)$.

{\em\color{red}Frequency-dependent bands.}---To clarify how self-energy renormalizes the eigenvalues, we introduce frequency-dependent bands. 
These bands are defined by the complex eigenvalues $E_n(\omega)$ of the non-Hermitian response Hamiltonian $H^{\rm Response}(\omega)$, satisfying
\begin{equation}
	H^{\rm Response}(\omega)|u_n^R(\omega)\rangle=E_n(\omega)|u_n^R(\omega)\rangle,
\end{equation}
where $|u_n^R(\omega)\rangle$ is the corresponding right eigenstate.
The real parts of these eigenvalues, $\Re [E_n(\omega)]$, are plotted as solid lines in Fig.~\ref{F2} (e) and \ref{F2} (f) for PBC and OBC, respectively. Their imaginary parts, $\Im [E_n(\omega)]$, are represented by the color scale, with redder hues indicating larger values.
The frequency-dependent bands also reveal two resonance peaks near \(\omega \sim -2.8\) and \(3.7\), which align with those found in the commutator norm \(C(\omega)\). 
This agreement confirms that both the spectra and eigenstates undergo the most significant renormalization at these frequencies, defining the fundamental energy scales of the interactions.

\begin{figure}[t] 
	\centering
	\includegraphics[width=1\columnwidth]{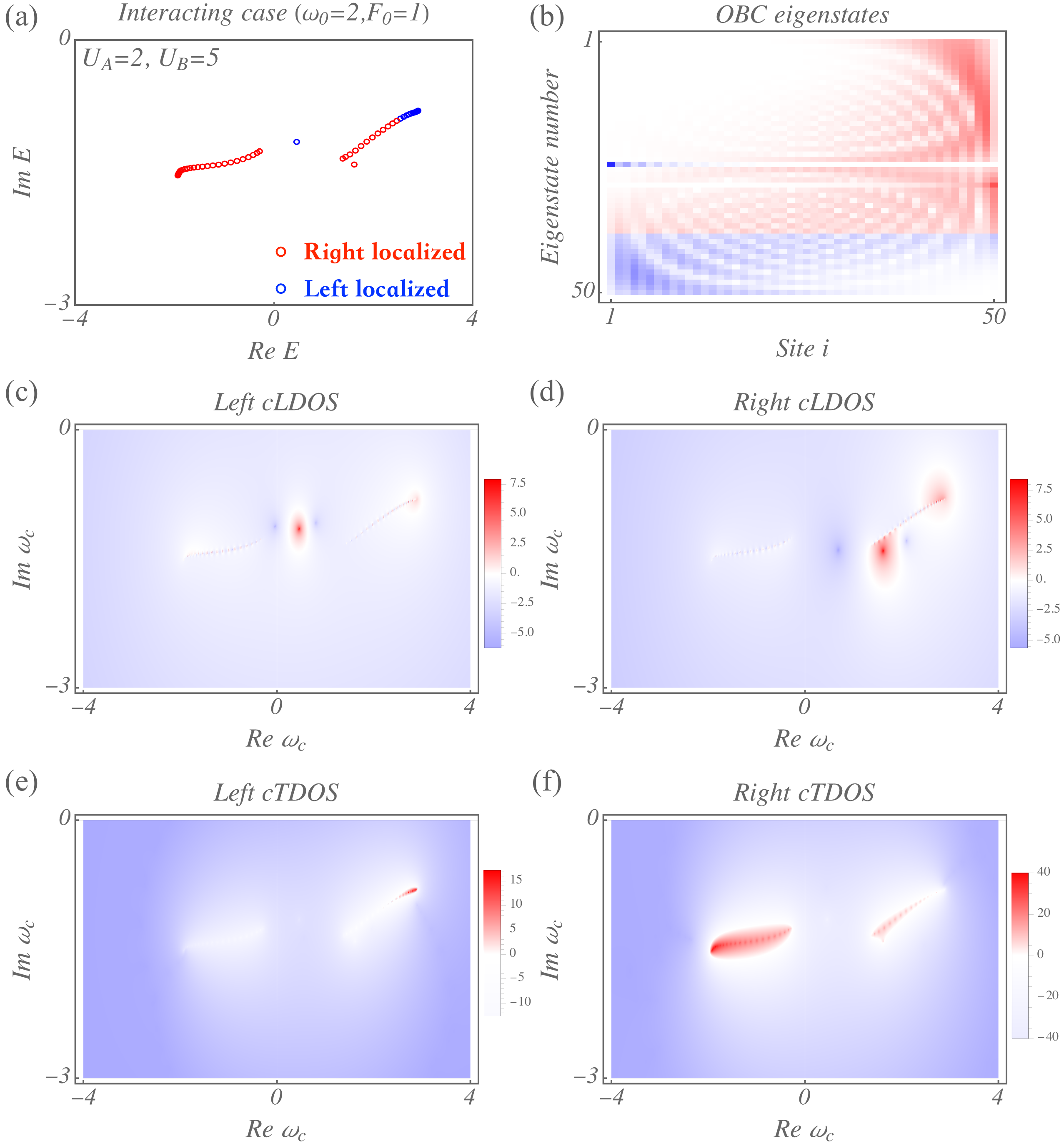}
	\caption{
		(a) Energy spectrum in the interacting case with parameters $\omega_0 = 2$, $U_A = 2$, and $U_B = 5$.  
		(b) Spatial distribution of eigenstates. 
		Red and blue colors indicate right- and left-localized states, respectively. 
		Localization behavior is determined based on the mean position of each eigenstate.  
		(c,d) Left and right complex frequency local density of states (cLDOS). 
		Resonance peaks in the complex frequency plane (shown in red) correspond to boundary-localized quasiparticles. 
		Only topological edge states exhibit dominant contributions, confirming their localized nature.  
		(e,f) Left and right complex frequency tunneling density of states. 
		Resonance peaks (red) reflect transmission quasiparticles in the complex plane. 
		These results indicate that skin modes are not localized but instead act as one-way transmission modes. }
	\label{F3}
\end{figure}

{\color{red}\em Complex frequency density of states.}---In this final section, we discuss how our method can be used to reveal complex frequency resonance peaks.
We take the case of $\omega_0=2$, $U_A=2$, and $U_B=5$ as an illustrative example. 
As shown in Fig.~\ref{F3} (a) and \ref{F3} (b), there are four distinct types of localization modes: left/right-localized topological edge states, which are represented by the blue/red isolated points in Fig.~\ref{F3} (a), and left/right-localized non-Hermitian skin modes, which are represented by the the blue/red quasi-continuum points in Fig.~\ref{F3} (a).
Here, the eigenstate index in Fig.~\ref{F3} (b) is ordered by the real part of their eigenvalues. 
The right and left localized eigenstates are defined by the mean position, $\sum_{m=1,...,2N} m |\langle m|u^{R}_{n}(\omega_{0})\rangle|^2$, i.e., a state is defined as left-localized if this value lies in $[1, N]$, and right-localized if it lies in $[N, 2N]$, where $N$ is the number of unit cells, and $|u^{R}_{n}(\omega_{0})\rangle$ is normalized.

To distinguish between these localized eigenstates, we first focus on local information near the boundary. 
This is achieved using the following CFF function~\cite{CFF}:
\begin{equation}
	G^{\rm CFF}(\omega_c \in \mathbb{C}) = \frac{1}{(\omega_c - \omega_0) + [\chi_{\rm steady}(\omega_0)]^{-1}}.
\end{equation}
Here, $\omega_c$ is a complex frequency that scans the entire complex plane, aiming to characterize the resonance peaks of quasiparticles. 
For instance, the left complex frequency local density of states (left cLDOS)~\cite{Prl125186,PraHenning} is defined as:
\begin{equation}
	{\rm cLDOS}_{\rm Left} = \sum_{i=1A,1B} \log |G^{\rm CFF}_{ii}(\omega_c)|.
\end{equation}
By evaluating this response function across the complex $\omega_c$ plane, we obtain the plot in Fig.~\ref{F3} (c). 
A comparison with Fig.~\ref{F3} (a) shows that only the left-localized topological edge state produces a pronounced complex frequency resonance peak (red color). 
Notably, the left-localized skin modes contribute negligibly to the ${\rm cLDOS}_{\rm Left}$~\cite{brody,Prl121kunst,djj}.
Similarly, the right cLDOS,
\begin{equation}
	{\rm cLDOS}_{\rm Right} = \sum_{i=NA,NB} \log |G^{\rm CFF}_{ii}(\omega_c)|,
\end{equation}
only captures the right-localized topological edge state, as shown by the red peak in Fig.~\ref{F3} (d). 
Again, the right-localized skin modes show no contribution. 
This result indicates that skin modes are not conventionally localized, as they do not contribute strongly to the boundary local density of states across the complex frequency plane.

To reveal the resonance peaks corresponding to the skin modes, we introduce a new concept: the complex frequency tunneling density of states (cTDOS). 
For example, the left cTDOS is defined as:
\begin{equation}
	{\rm cTDOS}_{\rm Left} = \sum_{\substack{i=1A,1B \ j=NA,NB}} \log |G^{\rm CFF}_{ij}(\omega_c)|.
\end{equation}
This quantity identifies which quasiparticles contribute to the tunneling process from the right boundary to the left boundary. 
As shown in Fig.~\ref{F3} (e), the left-localized skin modes now exhibit clear resonance peaks across the complex plane.
Conversely, the right cTDOS,
\begin{equation}
	{\rm cTDOS}_{\rm Right} = \sum_{\substack{i=NA,NB \ j=1A,1B}} \log |G^{\rm CFF}_{ij}(\omega_c)|,
\end{equation}
identifies the right-localized skin modes, as shown in Fig.~\ref{F3} (f).
These results further demonstrate that skin modes are indeed not localized modes, but are instead transmission modes, as their complex frequency resonance peaks appear only in the cTDOS and not in the cLDOS.

We note that this conclusion is specific to the 1D non-reciprocal skin effect. 
A corresponding study for reciprocal skin modes (e.g., in the 2D geometry-dependent non-Hermitian skin effect~\cite{gdse}) will be presented in a forthcoming paper.

{\color{red}\em Discussion and conclusion.}---In this work, we have systematically investigated the interplay between interactions and non-Hermitian topology in a time-reversal symmetry-broken SSH model with on-site Hubbard interactions. 
By employing the CFF method, we constructed an effective response Hamiltonian whose spectral and eigenstate properties reveal rich physical behavior beyond the non-interacting limit. 
Our key finding is that sublattice-asymmetric Hubbard interactions (\(U_A \neq U_B\)) can induce a point gap in the spectrum under PBC, accompanied by a pronounced non-Hermitian skin effect under OBC. 
This interaction-induced topological behavior is captured by the frequency-dependent commutator norm \(C(\omega_0)\), which quantifies the non-commutativity between the non-interacting Hamiltonian and the interaction-induced self-energy. 

Our results provide a general framework for studying interaction effects from a single-particle perspective in both quantum and classical systems. 
Even beyond the semiclassical approximation, the same mathematical structure, i.e., a frequency-dependent matrix of dimension $mN$, governs the response in full quantum systems, where $m$ is the number of internal degrees pf freedom.
The concepts of the commutator norm and frequency-dependent bands offer clear markers of the nontrivial role of interactions. 
This establishes a new paradigm for understanding strongly correlated systems, as the full single-particle Green’s function can be computed numerically through various methods, opening avenues for connecting effective non-Hermitian descriptions with many-body physics~\cite{yoshidaprb,PhysRevB.102.235151,PhysRevB.103.125145,PhysRevB.107.195149,PhysRevLett.133.076502,PhysRevB.100.115124,PhysRevLett.121.203001,PhysRevLett.123.123601,Prlshaokai,vvrx-mljg,cw1,Prldmft,PhysRevLett.133.086301,PhysRevB.102.081115,hao2025interactingmanybody,PhysRevLett.132.116503,PhysRevLett.129.180401,PhysRevB.101.235150,PhysRevB.102.064206,PhysRevA.99.011601,PhysRevResearch.1.033051,PhysRevLett.129.203401,PhysRevLett.130.123602,p5fv-c9dj,PhysRevB.108.075121,PhysRevResearch.5.L022046,wang2024w}.

Looking forward, our work opens several promising directions. 
The proposed experimental diagnostics, such as the complex frequency local and tunneling density of states measures, offer concrete strategies to distinguish between various types of localized modes in real experiments. 
It would be particularly interesting to explore the implications of such interaction-induced point gaps beyond the semiclassical approximation, where entanglement and quantum fluctuations may lead to even richer phenomena. 
Finally, our approach can be extended to higher dimensions and other symmetry classes, paving the way for a more comprehensive understanding of topology and interactions in both equilibrium and non-equilibrium settings.

Z. Yang, Z. Wang, J. Huang, and Z. Zheng were sponsored by the National Key R$\&$D Program of China (No. 2023YFA1407500) and the National Natural Science Foundation of China (No. 12322405). 
J. Hu was sponsored by the Ministry of Science and Technology (Grant No. 2022YFA1403901), National Natural Science Foundation of China (No. 12494594), and the New Cornerstone Investigator Program.

\section*{Appendix I: Derivation of the non-linear differential equation}
In this section, we provide a detailed derivation of the dynamical equation, i.e., Eq.~\ref{Eq4}. To begin, the dynamics of the entire system, including the external drive, are governed by the quantum master equation:
\begin{equation}
\frac{d}{dt}\hat{\rho}^{\boldsymbol{F}}(t)=-i[\hat{H}+\hat{H}^{\boldsymbol{F}}(t),\hat{\rho}^{\boldsymbol{F}}(t)]+\sum\limits^{2N}_{i=1}\kappa_{i}\hat{\mathcal{L}}_{i}[\hat{\rho}^{\boldsymbol{F}}(t)],
\end{equation}
where $\hat{\rho}^{\boldsymbol{F}}(t)$ denotes the density matrix operator at time $t$, and
\begin{equation}
\hat{H}^{\boldsymbol{F}}(t)=\sum^{2N}_{i=1}\big(\hat{\psi}^{\dag}_{i}F_{i}(t)+h.c.\big)
\end{equation}
represents the coupling to the harmonic driving fields with an arbitrary vector $\boldsymbol{F}$. Notably, in Eq.~\ref{Eq3} we have chosen $\boldsymbol{F}=\boldsymbol{F}_{j}$ to define the response function within the CFF method. Besides, $\kappa_{i}$ is the dissipation strength constants that quantify the respective damping rates for various dissipative processes. The Lindblad superoperators are defined as
\begin{equation}
\hat{\mathcal{L}}_{i}[\hat{\rho}^{\boldsymbol{F}}(t)]=\hat{\psi}_{i}\hat{\rho}^{\boldsymbol{F}}(t)\hat{\psi}^{\dag}_{i}-\frac{1}{2}\{\hat{\psi}^{\dag}_{i}\hat{\psi}_{i},\hat{\rho}^{\boldsymbol{F}}(t)\}.
\end{equation}
Then, we define the response to the external driving field for the bosonic field operator as $\boldsymbol{\psi}^{\boldsymbol{F}}(t)=\langle \hat{\boldsymbol{\psi}}(t)\rangle_{\boldsymbol{F}}=\mathrm{Tr}[\hat{\boldsymbol{\psi}}\hat{\rho}^{\boldsymbol{F}}(t)]$. Subsequently, according to the quantum master equation, the dynamical equation for the response is given by:

\begin{equation}
\begin{split}
&i \frac{d\langle \hat{\psi}_{i}(t)\rangle_{\boldsymbol{F}}}{dt}\\
&=i \mathrm{Tr}\big[\hat{\psi}_{i}\big(-i[\hat{H}+\hat{H}^{\boldsymbol{F}}(t),\hat{\rho}^{\boldsymbol{F}}(t)]+\sum\limits_{j}\kappa_{j}\hat{\mathcal{L}}_{j}[\hat{\rho}^{\boldsymbol{F}}(t)]\big)\big]\\
&=\mathrm{Tr}\big[\hat{\psi}_{i}[\hat{H}+\hat{H}^{\boldsymbol{F}}(t),\hat{\rho}^{\boldsymbol{F}}(t)]\big]+i\mathrm{Tr}\big[\hat{\psi}_{i}\sum\limits_{j}\kappa_{j}\hat{\mathcal{L}}_{j}[\hat{\rho}^{\boldsymbol{F}}(t)]\big].
\end{split}
\end{equation}
We then employ the identity:
\begin{equation}
\begin{split}
\mathrm{Tr}\big[\hat{\psi}_{i}[\hat{H}+\hat{H}^{\boldsymbol{F}}(t),\hat{\rho}^{\boldsymbol{F}}(t)]\big]=\mathrm{Tr}\big[[\hat{\psi}_{m},\hat{H}+\hat{H}^{\boldsymbol{F}}(t)]\hat{\rho}^{\boldsymbol{F}}(t)\big].
\end{split}
\end{equation}
Subsequently, utilizing the bosonic commutation relation $[\hat{\psi}_{i},\hat{\psi^{\dag}}_{j}]=\delta_{ij}$, we arrive at
\begin{equation}
[\hat{\psi}_{i},\hat{H}+\hat{H}^{\boldsymbol{F}}(t)]=\sum\limits_{j}t_{ij}\hat{\psi}_{j}+U_{i}\hat{\psi}^{\dag}_{i}\hat{\psi}_{i}\hat{\psi}_{i}+F_{i}(t).
\end{equation}
Then by applying the identity
\begin{equation}
\begin{split}
&\mathrm{Tr}\big[\hat{\psi}_{i}\big(\hat{\psi}_{j}\hat{\rho}^{\boldsymbol{F}}(t) \hat{\psi}^{\dag}_{j}-\frac{1}{2}\{\hat{\psi}^{\dag}_{j}\hat{\psi}_{j},\hat{\rho}^{\boldsymbol{F}}(t)\}\big)\big]\\
&=\mathrm{Tr}\big[\big(\hat{\psi}^{\dag}_{j}\hat{\psi}_{i}\hat{\psi}_{j}-\frac{1}{2}\{\hat{\psi}_{i},\hat{\psi}^{\dag}_{j}\hat{\psi}_{j}\}\big)\hat{\rho}^{\boldsymbol{F}}(t)\big]\\
&=\mathrm{Tr}\big[\frac{1}{2}\big([\hat{\psi}^{\dag}_{j},\hat{\psi}_{i}]\hat{\psi}_{j}+\hat{\psi}^{\dag}_{j}[\hat{\psi}_{i},\hat{\psi}_{j}]\big)\hat{\rho}^{\boldsymbol{F}}(t) \big]\\
&=-\frac{1}{2}\delta_{ij}\langle \hat{\psi}_{j}(t)\rangle_{\boldsymbol{F}},
\end{split}
\end{equation}
we derive
\begin{equation}
i\mathrm{Tr}\big[\hat{\psi}_{i}\sum\limits_{j}\kappa_{j}\hat{\mathcal{L}}_{j}[\hat{\rho}^{\boldsymbol{F}}(t)]\big]=-\frac{i}{2}\kappa_{i}\langle \hat{\psi}_{i}(t)\rangle_{\boldsymbol{F}}.
\end{equation}
Consequently, we establish
\begin{equation}
\begin{split}
&i\frac{d\langle \hat{\psi_{i}}(t)\rangle_{\boldsymbol{F}}}{dt}\\
&=\sum\limits_{j}(t_{ij}-\frac{i}{2}\kappa_{i}\delta_{ij})\langle\hat{\psi}_{j}(t) \rangle_{\boldsymbol{F}}+U_{i}\mathrm{Tr}\big[\hat{\psi}^{\dag}_{i}\hat{\psi}_{i}\hat{\psi}_{i}\hat{\rho}^{\boldsymbol{F}}(t)\big]+F_{i}(t).
\end{split}
\end{equation}
Notably, this exact dynamical equation has strong nonlinearity and cannot be closed for $\langle \hat{\psi_{i}}(t)\rangle_{\boldsymbol{F}}$ due to the interaction term.
In order to proceed, we adopt a semiclassical approximation in our analysis:
\begin{equation}
\begin{split}
\mathrm{Tr}\big[\hat{\psi}^{\dag}_{i}\hat{\psi}_{i}\hat{\psi}_{i}\hat{\rho}(t)\big]&\approx \langle\hat{\psi}^{\dag}_{i}(t)\rangle_{\boldsymbol{F}}\langle\hat{\psi}_{i}(t)\rangle_{\boldsymbol{F}}\langle\hat{\psi}_{i}(t)\rangle_{\boldsymbol{F}}\\
&\approx |\langle\hat{\psi}_{i}(t)\rangle_{\boldsymbol{F}}|^{2}\langle\hat{\psi}_{i}(t)\rangle_{\boldsymbol{F}}.
\end{split}
\end{equation} 
This approximation decouples the many-body operator correlation, leading to a closed form of the dynamical equation while preserving the non-linear effect.
Finally, if we introduce a non-linear diagonal matrix,
\begin{equation}
\mathcal{H}_{i}=\mathrm{diag}\big(U_{A}|\psi^{\boldsymbol{F}}_{1}(t)|^{2},U_{B}|\psi^{\boldsymbol{F}}_{2}(t)|^{2},\dots\big),
\end{equation}
and use $[\mathcal{H}_{0}]_{ij}=t_{ij}$ to represent the hopping matrix, and let $i\frac{\kappa_{i}}{2}=i\eta$ denote a uniform dissipation simulating the numerical convergence parameter, we can obtain the non-linear equation for the response
\begin{equation}
\big[i\frac{\partial}{\partial t}-(\mathcal{H}_{0}-i\eta)-\mathcal{H}_{i}\big]\boldsymbol{\psi}^{\boldsymbol{F}}(t)=\boldsymbol{F}(t).
\end{equation}
when we choose $\boldsymbol{F}(t)=\boldsymbol{F}_{j}e^{-i\omega_{0}t}\theta(t)$ with a real frequency $\omega_{0}\in\mathbb{R}$, then Eq.~\ref{Eq4} and Eq.~\ref{diaH} are achieved.

\section*{Appendix II: Perturbation analysis}

For a system in a steady state under external driving, the response function $\chi_{\rm s}$ (short for $\chi_{\rm steady}$) satisfies
\begin{equation}
	(\omega_0-\mathcal{H}_0+i\eta){\chi}_{\rm s}-U_{F_0}\circ {\chi}_{\rm s}^* \circ  {\chi}_{\rm s} \circ {\chi}_{\rm s} = I.
\end{equation}
Here $\omega_0$ is the driving frequency, and $U_{F_0}$ is a modified interaction matrix with elements that depend only on the row index:
\begin{equation}
	[U_{F_0}]_{ij}=\begin{cases}
		U_AF_0^2,& i=1,3,5,...\\
		U_BF_0^2, & i=2,4,6,...
	\end{cases}
\end{equation}
The symbol $\circ$ denotes the Hadamard product ($[A\circ B]{ij}=A_{ij}B_{ij}$), ${\chi}_{\rm s}^*$ is the complex conjugate of ${\chi}_{\rm s}$, and $I$ is the identity matrix.

We now treat the interaction term as a perturbation and expand ${\chi}_{\rm s}$ order-by-order:
(i) At zeroth order, ${\chi}_{\rm s}^{(0)}$ is the retarded Green's function for the non-interacting case:
\begin{equation}
	{\chi}_{\rm s}^{(0)}=\frac{1}{\omega_0-\mathcal{H}_0+i\eta};
\end{equation}
(ii) The $ m$th-order term obeys the recurrence relation:
\begin{equation}
	{\chi}_{\rm s}^{(m \geq1)}={\chi}_{\rm s}^{(0)}\left[\sum_{n_1+n_2+n_3=m}U_{F_0}\circ {\chi}_{\rm s}^{*,(n_1)}\circ {\chi}_{\rm s}^{(n_2)}\circ {\chi}_{\rm s}^{(n_3)}\right]. 
\end{equation}
This framework allows for the calculation of the response function to any arbitrary order.

This analysis reveals an interesting duality in our model: the response function depends solely on the product $U_\alpha F_0^2$. 
This implies a duality between a weakly driven system with strong interactions and a strongly driven system with weak interactions. 

\bibliography{ref}
\bibliographystyle{apsrev4-1}
	
\end{document}